\documentclass[conference]{IEEEtran}
\IEEEoverridecommandlockouts
\usepackage{cite}
\usepackage{amsmath,amssymb,amsfonts}
\usepackage{algorithm}
\usepackage{algpseudocode}
\algrenewcommand\algorithmicrequire{\textbf{Initialize:}}
\usepackage{graphicx}
\usepackage{textcomp}
\usepackage{xcolor}
\def\BibTeX{{\rm B\kern-.05em{\sc i\kern-.025em b}\kern-.08em
    T\kern-.1667em\lower.7ex\hbox{E}\kern-.125emX}}
\begin{document}

\title{
Schwarz Information Criterion Aided MAB for Resource Allocation in Dynamic LoRa System\\
{}
}


\author{Ryota Ariyoshi\IEEEauthorrefmark{1},
Aohan Li\IEEEauthorrefmark{1},
Mikio Hasegawa\IEEEauthorrefmark{2}, 
Miao Pan\IEEEauthorrefmark{3},
Tomoaki Ohtsuki\IEEEauthorrefmark{4},
and Zhu Han\IEEEauthorrefmark{3}
\\
\IEEEauthorrefmark{1}Department of Computer and Network Engineering, The University of Electro-Communications, Tokyo, Japan\\
\IEEEauthorrefmark{2}Department of Electrical Engineering, Tokyo University of Science, Tokyo, Japan\\
\IEEEauthorrefmark{3}Department of Electrical and Computer Engineering, University of Houston, Houston, TX, 77204, USA\\
\IEEEauthorrefmark{4}Department of Information and Computer Science, Keio University, Yokohama, Japan\\
}


\maketitle

\begin{abstract}

This paper proposes a lightweight distributed learning method for transmission parameter selection in Long Range (LoRa) networks that can adapt to dynamic communication environments. In the proposed method, each LoRa End Device (ED) employs the Upper Confidence Bound (UCB)1-tuned algorithm to select transmission parameters including channel, transmission power, and bandwidth. The transmission parameters are selected based on the ACKnowledgment (ACK) feedback returned from the gateway after each transmission and the corresponding transmission energy consumption. Hence, it enables devices to simultaneously optimize transmission success rate and energy efficiency in a fully distributed manner.  
However, although UCB1-tuned based method is effective under stationary conditions, it suffers from slow adaptation in dynamic environments due to its strong reliance on historical observations. 
To address this limitation, we integrate the Schwarz Information Criterion (SIC) to our proposed method. SIC is adopted because it enables low-cost detection of changes in the communication environment, making it suitable for implementation on resource-constrained LoRa EDs. When a change is detected by SIC, the learning history of UCB1-tuned is reset, allowing rapid re-learning under the new conditions.
Experimental results using real LoRa devices demonstrate that the proposed method achieves superior transmission success rate, energy efficiency, and adaptability compared with the conventional UCB1-tuned algorithm without SIC.

\end{abstract}

\begin{IEEEkeywords}
Dynamic LoRa networks, Decentralized Resource Allocation, Energy Efficiency, Multi-Armed Bandit, Schwarz Information Criterion 
\end{IEEEkeywords}

\section{Introduction}
\label{Introduction}

With the rapid proliferation of the Internet of Things (IoT) devices in recent years, it has become increasingly common for a massive number of terminals to transmit data simultaneously~\cite{b1}. 
This explosive growth has intensified channel contention in Low-Power Wide-Area Networks (LPWANs), particularly in Long Range (LoRa) networks, which stand out as a prominent example of LPWAN technology~\cite{b3,b5}.
To use limited wireless resources efficiently, each LoRa device need to optimize its transmission parameters in response to its dynamically changing communication environment.

To address these challenges, previous studies have explored centralized approaches, in which a network management server controls the transmission parameters of all devices. However, centralized methods suffer from scalability limitations, since the processing load is concentrated on the network server~\cite{b6,b7}. Moreover, the gateway must inform each End Device (ED) of the transmission parameters before every transmission. This increases energy consumption and latency because the EDs must remain in receive mode~\cite{b1}. In contrast, decentralized approaches allow each device to autonomously determine its transmission parameters without waiting for gateway instructions. As a result, they not only provide superior scalability but also reduce energy consumption and communication resources compared with centralized methods.

One representative decentralized technique used in distributed resource allocation is the Multi-Armed Bandit (MAB) algorithms~\cite{b8,b9,b10,b11,b12,b13}. MAB is a reinforcement learning method that balances exploration and exploitation to learn the optimal choice, and it has been well applied to transmission parameter optimization. In particular, Upper Confidence Bound (UCB)1-tuned algorithm, which considers the variance of rewards for each option, has been widely used due to its enhanced exploration efficiency. Such decentralized online learning mechanisms can be regarded as a lightweight form of AI-driven networking, where EDs autonomously learn and adapt their transmission policies based on local observations. 
However, most MAB based distributed transmission parameter selection methods, including UCB1-tuned, rely heavily on historical observations. This dependence results in delayed adaptation when the communication environment changes abruptly, as these algorithms lack mechanisms to automatically discard past learning history~\cite{b14}. Consequently, performance degradation in dynamic environments becomes unavoidable. 

To overcome the limitation discussed above, this paper incorporates the Schwarz Information Criterion (SIC) into a decentralized UCB1-tuned learning process for transmission-parameter selection in dynamic LoRa networks.
In the proposed method, SIC detects variations in channel statistics and resets the learning process, enabling rapid and fully distributed adaptation in non-stationary environments~\cite{b15}. 
By integrating SIC with the UCB1-tuned learning process, the proposed method enables practical implementation on memory-constrained IoT devices while maintaining a high transmission success rate and energy efficiency under dynamic communication conditions.
The main contributions of this paper are summarized as follows:
\begin{itemize}
    \item
    We propose a lightweight distributed learning based transmission parameter selection method. 
    In our proposed method, only a small set of statistics need to be stored, and requires neither complex matrix
operations nor centralized control. Consequently, it incurs
low computational and memory overhead, making it
well suited for implementation on memory-constrained
IoT devices. 
    \item 
    Our proposed method can adapt well to dynamically changing wireless environments by incorporating the SIC based change-detection principle.
    Additionally, energy efficiency is well considered in our proposed method by designing the relevant aspects of the learning method, i.e., the reward function.
    \item We implement the proposed method on real hardware and conduct extensive evaluations in a dynamic environment to validate its effectiveness. Experimental results demonstrate that our proposed method improves transmission success rate, energy efficiency, and adaptability compared with the conventional UCB1-tuned algorithm in a dynamic LoRa network.
\end{itemize}

This paper is structured as follows. Section \ref{System Model} describes the system model and problem formulation. Section \ref{Proposed Method} presents the proposed method. Section \ref{Performance Evaluation} provides the implementation and performance evaluation. Finally, Section \ref{Conclusion} concludes the paper.

\section{System Model}
\label{System Model}
\begin{figure}[t]
\centering
\includegraphics[width=7.3cm]{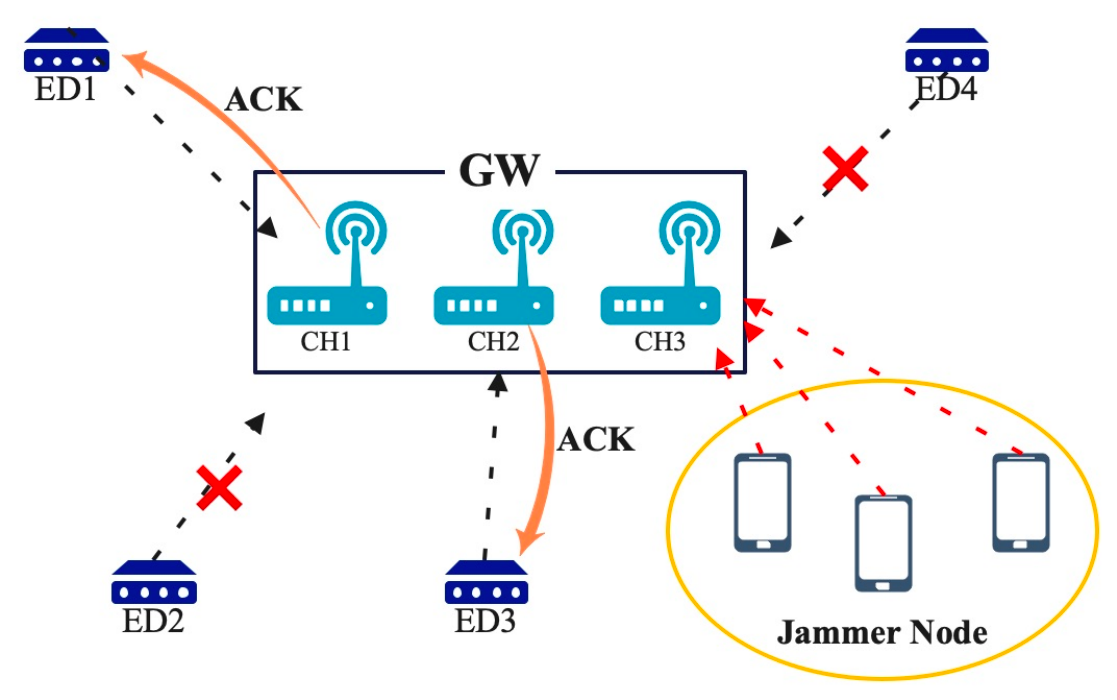}
\caption{System Model}
\label{fig:s1}
\end{figure}

The LoRa system considered in this study is illustrated in Fig.~\ref{fig:s1}, consisting of a single GateWay (GW), multiple EDs, and an external interference source (jammer node). In the LoRa system, multiple channels are available in the network, and each ED selects one of them to transmit data to the GW. However, when the jammer occupies or blocks a specific channel, transmissions over that channel are more likely to experience interference. Therefore, we assume a dynamic communication environment where the jammer intermittently affects part of the frequency spectrum, causing certain channels to transition from an available state to an unavailable state during operation. Each ED needs to adaptively select transmission parameters that achieve the highest energy efficiency under varying channel conditions.

Let $G$ be the number of LoRa EDs, $M$ be the number of available channels, $P$ be the number of transmission power (TP) levels, and $H$ be the number of bandwidth (BW) options. The set of available channels is denoted as 
$C = \{c_{1}, c_{2}, \ldots, c_{m}, \ldots, c_{M}\}$, 
the set of TP levels as 
$U = \{u_{1}, u_{2}, \ldots, u_{p}, \ldots, u_{P}\}$, 
and the set of BW options as 
$B = \{b_{1}, b_{2}, \ldots, b_{h}, \ldots, b_{H}\}$. 
Let $K = \{k_{1}, k_{2}, \ldots, k_{m \times p \times h}, \ldots, k_{M \times P \times H}\}$ be the set of all possible combinations of channel, TP, and BW.
Each ED selects one parameter set from 
$K$ before transmitting data based on our proposed method.

Each ED transmits data at fixed intervals. Prior to transmission, it performs carrier sensing on the selected channel. If the channel is sensed as available, ED proceeds to transmit using the selected combination of channel, TP, and BW. If the transmission is successful, the GW sends an Acknowledgment (ACK) feedback to the ED, and the ED receives a positive reward. If no ACK is received, the transmission is considered failed, and the result is reflected in subsequent learning. Parameter selection and learning processes are executed in a fully distributed manner on each ED. Based on ACK feedback and energy consumption information associated with the chosen parameters, each ED updates its learning policy to maximize both success rate of the transmission and energy efficiency.

Furthermore, each ED records the presence or absence of ACKs for each transmission and accumulates this as a transmission history. In this study, we introduce a mechanism that statistically detects changes in the communication environment based on this history using the SIC. When a change is detected, the past learning history is discarded, and the learning process is reinitialized to enable quick adaptation to the new environment. 
Each ED maintains $M$ binary observation sequences of ACK receptions, and the $m$-th sequence corresponds to $c_m$ is represented as follows:
\begin{equation}
    \Omega_{c_m} = \{\, s_i \mid s_i \in \{0, 1\},\, i = 1, 2, \ldots, l \,\},
\end{equation}
where $l$ denotes the length of the $\Omega_{c_m}$, $s_i = 1$ denotes a successful ACK reception, and $s_i = 0$ denotes a failure.
Let $l_{\max}$ denote the maximum length of each binary observation sequence. Historical data prior to $l_{\max}$ are discarded to maintain low memory overhead and real-time performance.
To analyze temporal variations in communication performance, this sequence is divided into sliding windows of length $W$, shifted by $F$ steps. 
Given the total sequence length $l$, the number of windows $D$ is expressed as:
\begin{equation}
D = \lfloor\frac{l + F - W}{F}\rfloor.
\end{equation}
Let $x_d$ denote the number of ACK successes in window $d \in \{1, 2, \ldots, D\}$,
and let $p_d$ represent the average ACK success probability within that window.
These statistics are then used to evaluate the temporal variations in transmission performance and to detect environmental changes.

The energy consumption model of a LoRa ED used in this paper models the energy consumption
during data communication in active mode, which can be calculated below \cite{b6}. 

    \begin{equation}
       E_{Active} = E_{WU} + E_{\text{proc}} + E_{\text{ToA}} + E_{\text{R}},
    \end{equation}   
    where \(E_{\text{WU}}\) represents the energy consumption during device wake-up, \(E_{\text{proc}}\) represents the energy consumption for transmission parameter selection by the microcontroller, \(E_{\text{ToA}}\) represents the energy consumption during data transmission, and \(E_{\text{R}}\) represents the energy consumption during the reception. The values of $E_{\text{WU}}$, \(E_{\text{proc}}\), and \(E_{\text{R}}\) depend on the specifications of the modules used in the device.
    \(E_{\text{ToA}}\) can be expressed as follows:
    \begin{equation}
       E_{\text{ToA}} = (P_{\text{MCU}} + P_{\text{ToA}}) \cdot T_{\text{ToA}}
       \label{EToA},
    \end{equation}  
    where \(P_{\text{MCU}}\) is the power consumption due to the activation of the microcontroller, \(P_{\text{ToA}}\) is the power consumption during data transmission, which is determined by the selected TP.
    \(T_{\text{ToA}}\) is the total transmission duration, which can be calculated as follows:
    \begin{equation}
      T_{\text{ToA}} = T_{\text{Preamble}} + T_{\text{Payload}},
    \end{equation}
    where \(T_{\text{Preamble}}\) represents the duration required to transmit the preamble, and \(T_{\text{Payload}}\) represents the duration required to transmit the data payload. \(T_{\text{Preamble}}\) and \(T_{\text{Payload}}\) can be expressed as follows:
    \begin{equation}
      T_{\text{Preamble}} = (4.25 + N_{\text{P}}) \cdot T_{\text{Symbol}},
    \end{equation}
    \begin{equation}
      T_{\text{Payload}} = N_{\text{Payload}} \cdot T_{\text{Symbol}},
    \end{equation}
    where \(N_{\text{P}}\) is the number of preamble symbols, and \(N_{\text{Payload}}\) is the number of payload symbols.
    \(T_{\text{Symbol}}\) is the symbol duration, which can be calculated as follows:
    \begin{equation}
      T_{\text{Symbol}} = \frac{2^{SF}}{BW},
    \end{equation}
    where spreading factor (\(SF\)) and \(BW\) are the used SF and BW when transmitting symbols.

In this study, the transmission success rate of a parameter set $k_i$ at time $t$ is defined as
\begin{equation}
X_{k_i}(t) = \frac{r_{k_i}(t)}{N_{k_i}(t)},
\end{equation}
where $r_{k_i}(t)$ is the cumulative number of successful transmissions and $N_{k_i}(t)$ is the cumulative number of times the parameter set $k_i$ has been selected up to time $t$. This value reflects the probability of successful transmission based on the past history of each parameter set.  

Based on this definition, the energy efficiency (EE) of a parameter set $k_i$ at time $t$ is expressed as

\begin{equation}
EE_{k_i}(t) = 
\frac{\text{Payload}_{k_i}(t) \times X_{k_i}(t)}{E_{\text{Active}}},
\label{eq:EE}
\end{equation}
where $\text{Payload}_{k_i}(t)$ [bit] denotes the payload size associated with the transmission parameter combination $k_i$ at time $t$.
In other words, $EE_{k_i}(t)$ indicates the number of successfully transmitted bits per unit of consumed energy (bit/J).

The objective of this study is to maximize the cumulative energy efficiency of all EDs by optimally selecting the channel, TP, and BW under dynamic communication environments. This optimization problem is formulated as
\begin{equation}
\max_{k_i \in K} \sum_{g=1}^{G} \sum_{t=1}^{T} EE_{k_i}^{(g)}(t),
\label{eq:objective_allEDs}
\end{equation}
where $EE_{k_i}^{(g)}(t)$ represents the instantaneous energy efficiency of device $g$ 
when it selects parameter combination $k_i$ at time $t$, and $T$ denotes the total number of transmissions.

\section{Proposed Method}
\label{Proposed Method}
This paper proposes a distributed reinforcement learning method that integrates the UCB1-tuned algorithm with SIC to maximize energy efficiency in dynamic LoRa networks. Each LoRa ED autonomously selects combination of channel, TP, and BW based on local observations. In the proposed method, the reward used in the UCB1-tuned algorithm is defined as the number of successfully transmitted bits per unit of consumed energy (bit/J), enabling simultaneous optimization of transmission reliability and energy efficiency. Additionally, when a significant environmental change is detected by SIC, the learning history is reset to allow rapid re-adaptation to new communication conditions.
This section presents the proposed method, which first introduces the concept of SIC-based change detection, followed by the integration of this mechanism into the distributed UCB1-tuned framework for adaptive transmission-parameter selection in dynamic LoRa networks.

\subsection{Schwarz Information Criterion}

To adapt to the non-stationarity of the LoRa communication environment, this paper introduces a statistical change detection mechanism based on SIC~\cite{b14}. Although SIC is generally employed as an information-theoretic criterion for model selection, it is applied here to quantitatively determine whether the statistical characteristics of the communication environment have changed. Since SIC evaluates the tradeoff between model fit and complexity, it can be used to compare statistical models that assume different ACK success probabilities before and after a potential change point. A significant increase in SIC indicates that a model assuming different success probabilities fits the observed ACK sequence better than a single-probability model, thereby signaling a change in the communication environment.



In the SIC, two hypotheses are considered:  
(a) the success probability remains constant across all windows ($H_0$), and  
(b) the success probability changes at an unknown point ($H_1$).

(a) Null Hypothesis (No Change) $H_0$:  
All windows share the same ACK success probability $p$, i.e., $p_1 = p_2 = \cdots = p_D = p$.  
The SIC under $H_0$ is calculated as
\begin{align}
\mathrm{SIC}(D) &= \log D - 2 \sum_{d=1}^{D} \log \binom{W}{x_d} \nonumber \\
&\quad - 2(Y - X)\log\left(\frac{Y - X}{Y}\right)
 - 2X\log\left(\frac{X}{Y}\right),
 \label{eq:sic_H0}
\end{align}
where $X = \sum_{d=1}^{D} x_d$ is the total number of ACK successes and $Y = D W$ is the total number of transmission attempts.

(b) Alternative Hypothesis (Change Exists) $H_1$:  
There exists a split point $j$ $(1 \le j < D)$ such that  
$p_1 = \cdots = p_j \neq p_{j+1} = \cdots = p_D$.  
The SIC under $H_1$ is expressed as
\begin{align}
\mathrm{SIC}(j) &= 2\log D - 2 \sum_{d=1}^{D} \log \binom{W}{x_d} \nonumber \\
&\quad - 2(Y_j - X_j)\log\left(\frac{Y_j - X_j}{Y_j}\right)
 - 2X_j\log\left(\frac{X_j}{Y_j}\right) \nonumber \\
&\quad - 2(Y'_j - X'_j)\log\left(\frac{Y'_j - X'_j}{Y'_j}\right)
 - 2X'_j\log\left(\frac{X'_j}{Y'_j}\right),
 \label{eq:sic_H1}
\end{align}
where $X_j = \sum_{d=1}^{j} x_d$, $Y_j = jW$, $X'_j = X - X_j$, and $Y'_j = Y - Y_j$.  
The split point $j$ represents a potential boundary in the observation sequence at which the ACK success probability may change. In other words, SIC compares the likelihoods of two statistical models: one assumes a constant ACK success probability, while the other assumes different probabilities before and after point $j$, to determine whether a change in the communication environment has occurred.

An environmental change is detected when the following condition is satisfied:
\begin{equation}
\mathrm{SIC}(D) - \min_{1 \le j \le D-1} \mathrm{SIC}(j) > \theta,
\label{eq:sic_threshold}
\end{equation}
where $\theta$ is an empirically determined detection threshold.
Smaller $\theta$ adapts faster but risks false resets; larger $\theta$ is steadier but slower to respond.
When this condition holds at time $t$, the ED resets its learning history for all parameter combinations $k_i \in K$ in the UCB1-tuned algorithm, including cumulative reward $R_{k_i}(t)$, selection count $N_{k_i}(t)$, variance estimate $V_{k_i}(t)$, UCB score $P_{k_i}(t)$ to zero.
In addition, the ACK sequences $\Omega_{c_m}$ for for all $c_m\in\mathcal{C}$ are reset to  empty set. This enables each ED to promptly adapt to new communication environments without being affected by learning histories.


\subsection{SIC-based Transmission Parameter Selection}
The overall procedure of the proposed method, which is independently executed by each ED, is summarized in Algorithm 1. Each ED updates its transmission policy using ACK feedback and energy consumption information. The algorithm balances exploration and exploitation through the UCB1-tuned strategy and maintains adaptability to environmental changes through statistical change detection using the SIC.

\begin{algorithm}[t]
\small
\caption{Proposed Method}
\label{alg:proposed}
\begin{algorithmic}[1]
\Require $t = 0$, $P_{k_i}(t) = 0$, $R_{k_i}(t) = 0$, $N_{k_i}(t) = 0$, $V_{k_i}(t) = 0$ for all $k_i \in K$, $\Omega_{c_m} \leftarrow \emptyset$
        for all $c_m\in\mathcal{C}$
\ForAll{parameter combinations $k_i \in K$}
    \State Transmit using $k_i$ and observe ACK $s_i \in \{0, 1\}$
    \State Append $s_i$ to the corresponding history $\Omega_{c_m}$
     \State $t \leftarrow t + 1$
\EndFor

\While{$t < T$}
    \State Select $k^* \leftarrow \textsc{UCB1-tuned}\bigl(t, P_{k_i}(t), R_{k_i}(t), N_{k_i}(t), V_{k_i}(t)\bigr)$
    \State Transmit using $k^*$ and observe ACK $s_t \in \{0, 1\}$
    \State Measure transmission energy consumption $E_{\mathrm{ToA}}$ using ~(\ref{EToA})
    \If{$s_t = 1$}
        \State $\Delta R \leftarrow \dfrac{Payload_{k^*}(t)}{E_{\mathrm{ToA}}}$
    \Else
        \State $\Delta R \leftarrow 0$
    \EndIf
    \State Update $R_{k^*}(t)$, $N_{k^*}(t)$, and $V_{k^*}(t)$ based on the reward $\Delta R$
    \State Compute $\mathrm{SIC}(D)$ and $\mathrm{SIC}(j)$ using ~(\ref{eq:sic_H0})--(\ref{eq:sic_H1}) based on the $\Omega_{c_m}$ that corresponds to the selected channel $c_m$
    \If{$\mathrm{SIC}(D) - \min_j \mathrm{SIC}(j) > \theta$}
        \State Reset all $R_{k_i}(t)$, $N_{k_i}(t)$, $V_{k_i}(t)$, and $P_{k_i}(t)$ to zero
        \State Reset history $\Omega_{c_m} \leftarrow \emptyset$
        for all $c_m\in\mathcal{C}$
    \EndIf
    \State $t \leftarrow t + 1$
\EndWhile

\Function{UCB1-tuned}{$t, P_{k_i}(t), R_{k_i}(t), N_{k_i}(t), V_{k_i}(t)$}
    \ForAll{$k_i \in K$}
        \State Calculate $P_{k_i}(t)$ using ~(\ref{eq:ucb})
    \EndFor
    \State $k^* \leftarrow \arg\max_{k_i \in K} P_{k_i}(t)$
    \Return $k^*$
\EndFunction

\end{algorithmic}
\end{algorithm}

The algorithm operates as follows. Initially (lines~1--5), each ED initializes all learning variables to zero and performs one transmission for each parameter combination to collect the initial ACK feedback. 
During the main loop (lines~6--22), each ED selects the parameter combination $k^*$ that maximizes the UCB score, balancing exploration and exploitation by considering both the empirical mean and the variance of observed rewards. 
For each parameter combination $k_i$, the UCB score is computed as
\begin{equation}
    P_{k_i}(t) = \frac{R_{k_i}(t)}{N_{k_i}(t)} + 
    \sqrt{\frac{\ln t}{N_{k_i}(t)} \cdot 
    \min\!\left(\frac{1}{4}, V_{k_i}(t)\right)}.
    \label{eq:ucb}
\end{equation}
The first term represents the empirical mean reward, and the second term promotes exploration of under-selected actions. 
The parameter combination with the largest $P_{k_i}(t)$ is selected for the next transmission (line~7).
After each transmission (lines~8--14), the ED observes the ACK result and calculates reward. 
If the ACK result $s_t=1$, the reward is calculated as
\begin{equation}
    \Delta R = \frac{\text{Payload}_{k^*}(t)}{E_{\text{ToA}}},
    \label{eq:reward}
\end{equation}
corresponding to a successful transmission; otherwise, $\Delta R=0$ for a failed one. 
Then, $R_{k^*}(t)$, $N_{k^*}(t)$, and $V_{k^*}(t)$ are updated based on the obtained reward and the observed ACK information (line~15).
Next, SIC is applied to the ACK reception history $\Omega_{c_m}$ to detect potential environmental changes (lines~16--20). 
When the threshold condition defined in Eq. (\ref{eq:sic_threshold}) is satisfied, the algorithm resets all learning variables and the ACK history to remove outdated information. The process continues until the maximum transmission count is reached. The subroutine UCB1-TUNED (lines~23--28) computes the UCB score for each $k_i$ using the current statistics and returns the combination $k^*$ with the highest score~\cite{b17}.

\section{Performance Evaluation}
\label{Performance Evaluation}
\begin{table}[t]
\caption{Experimental Parameter Settings}
\begin{center}
\begin{tabular}{|c|c|}
\hline
\textbf{Parameter} & \textbf{Value} \\
\hline
Number of EDs & 30 \\
\hline
channel & 920.7, 921.1 MHz (250 kHz), \\
              & 921.4, 921.6, 921.8 MHz (125 kHz) \\
\hline
TP & -3, 1, 5, 9, 13 dBm \\
\hline
SF & 7 \\
\hline
BW & 125, 250 kHz \\
\hline
Transmission Interval & 15 seconds \\
\hline
Retransmission Count & 0 \\
\hline
Number of Transmissions & 1000 \\
\hline
Payload Length & 50 bytes \\
\hline
Startup Energy ($E_{\mathrm{WU}}$) & $56.1 \times T_{\mathrm{WU}}$ [mWh] \\
\hline
Processing Energy ($E_{\mathrm{proc}}$) & $85.8 \times T_{\mathrm{proc}}$ [mWh] \\
\hline
Reception Energy ($E_{\mathrm{R}}$) & $66 \times T_{\mathrm{R}}$ [mWh] \\
\hline
MCU Power ($P_{\mathrm{MCU}}$) & 29.7 [mW] \\
\hline
Preamble Length ($N_{\mathrm{P}}$) & 8 symbols \\
\hline
Sliding Window Length ($W$) & 10 \\
\hline
Window Shift Step ($F$) & 5 \\
\hline
ACK Sequence Length ($l_{\max}$) & 25 \\
\hline
Threshold ($\theta$) & 20 \\
\hline
\end{tabular}
\label{tab:param_settings}
\end{center}
\end{table}
\begin{figure}[!t]
\centerline{\includegraphics[width=75mm]{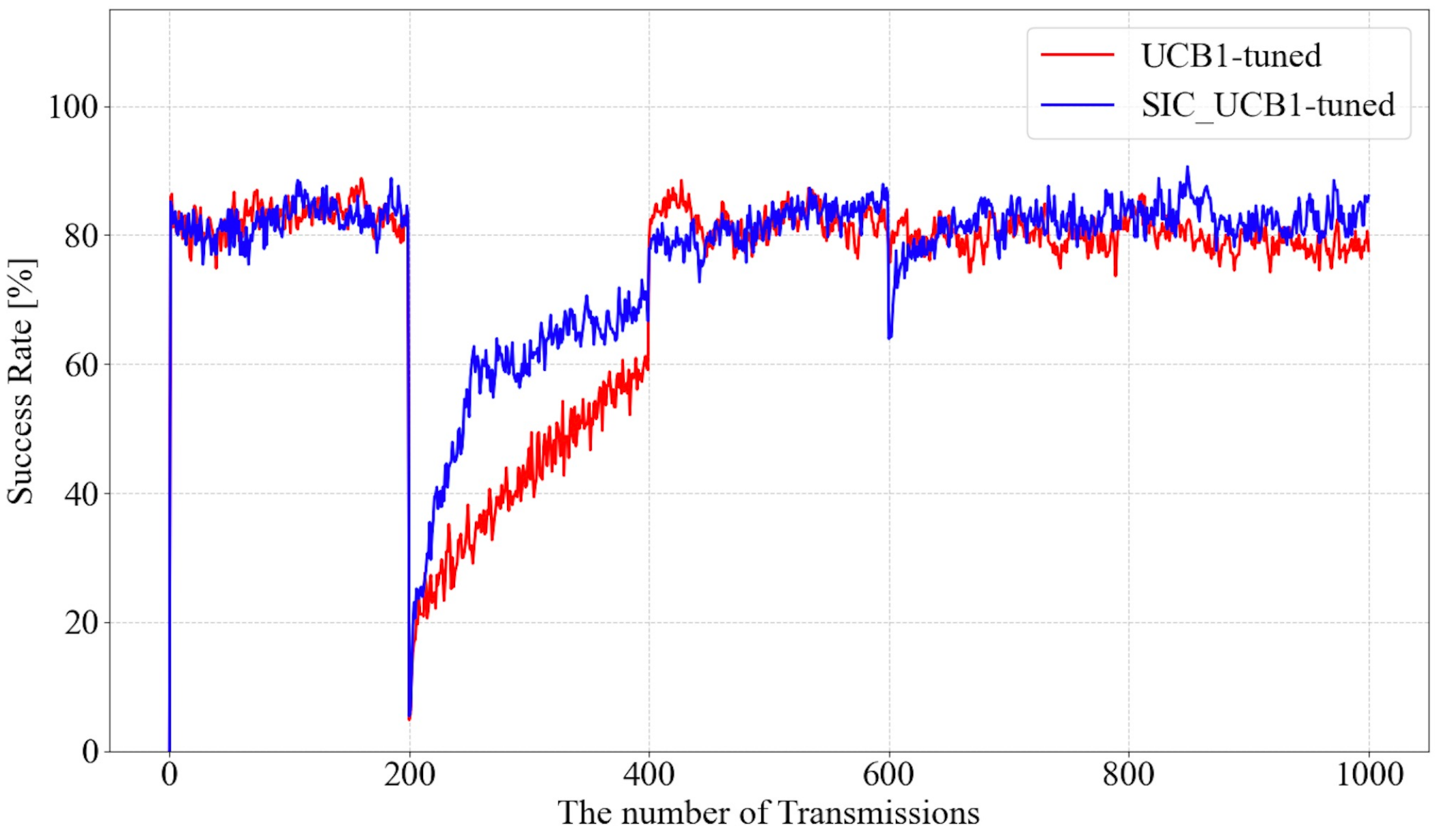}}
\caption{Transmission success rate versus the number of transmissions.}
\label{SIC_SR}
\end{figure}
To validate the effectiveness of the proposed method under dynamically changing communication environments, we conducted comparative experiments between the proposed SIC-based method and the conventional UCB1-tuned algorithm without SIC. In both methods, each ED autonomously selects transmission parameters (channel, TP, BW) in a distributed manner based on UCB1-tuned. The key distinction of the proposed method lies in its ability to statistically detect environmental changes using SIC and reset the learning process accordingly.

In this experiment, the communication environment was dynamically varied in three phases to emulate realistic non-stationary channel conditions. 
During the transmission intervals of 1--200, 401--600, and 801--1000, all five channels (920.7, 921.1, 921.4, 921.6, and 921.8 MHz) were available for successful reception, representing a stable communication environment. 
During 201--400 transmissions, the 250 kHz channels (920.7 MHz and 921.1 MHz) were intentionally disabled to simulate strong interference or physical blockage in a specific frequency band. 
Subsequently, during 601--800 transmissions, the 125 kHz channels (921.4 MHz and 921.6 MHz) were disabled to emulate a change in the communication environment affecting another frequency band. 
Under this dynamically varying environment, the proposed method and the conventional UCB1-tuned method were compared in terms of transmission success rate and energy efficiency.
The experiment was repeated ten times, and the average values were used to evaluate the success rate and energy efficiency. 
The detailed experimental parameters are summarized in Table~\ref{tab:param_settings}.

\subsection{Success rate}

\begin{figure}[!t]
\centerline{\includegraphics[width=92mm]{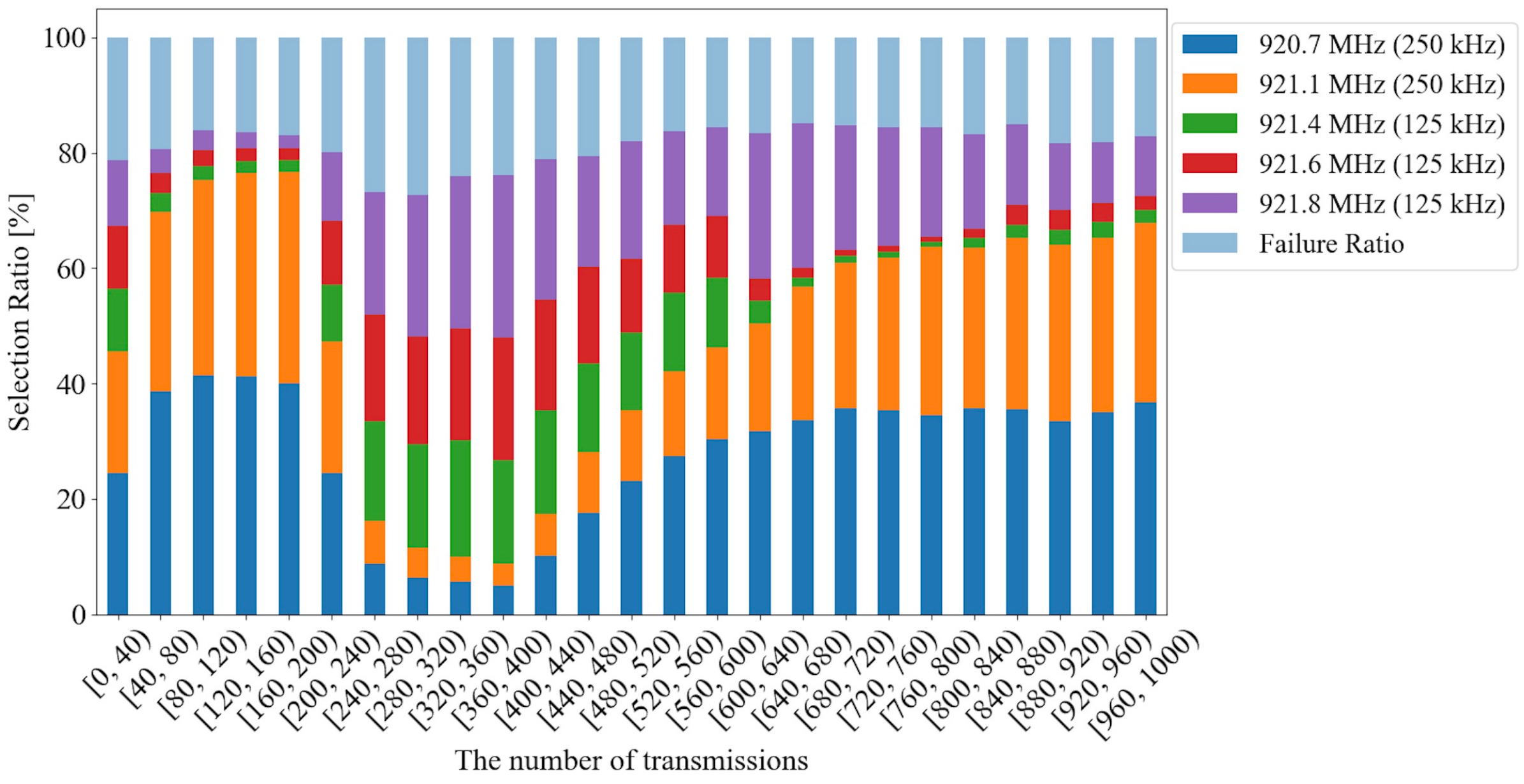}}
\caption{Selection ratio (proposed method)}
\label{SIC_0926}
\end{figure}
\begin{figure}[!t]
\centerline{\includegraphics[width=92mm]{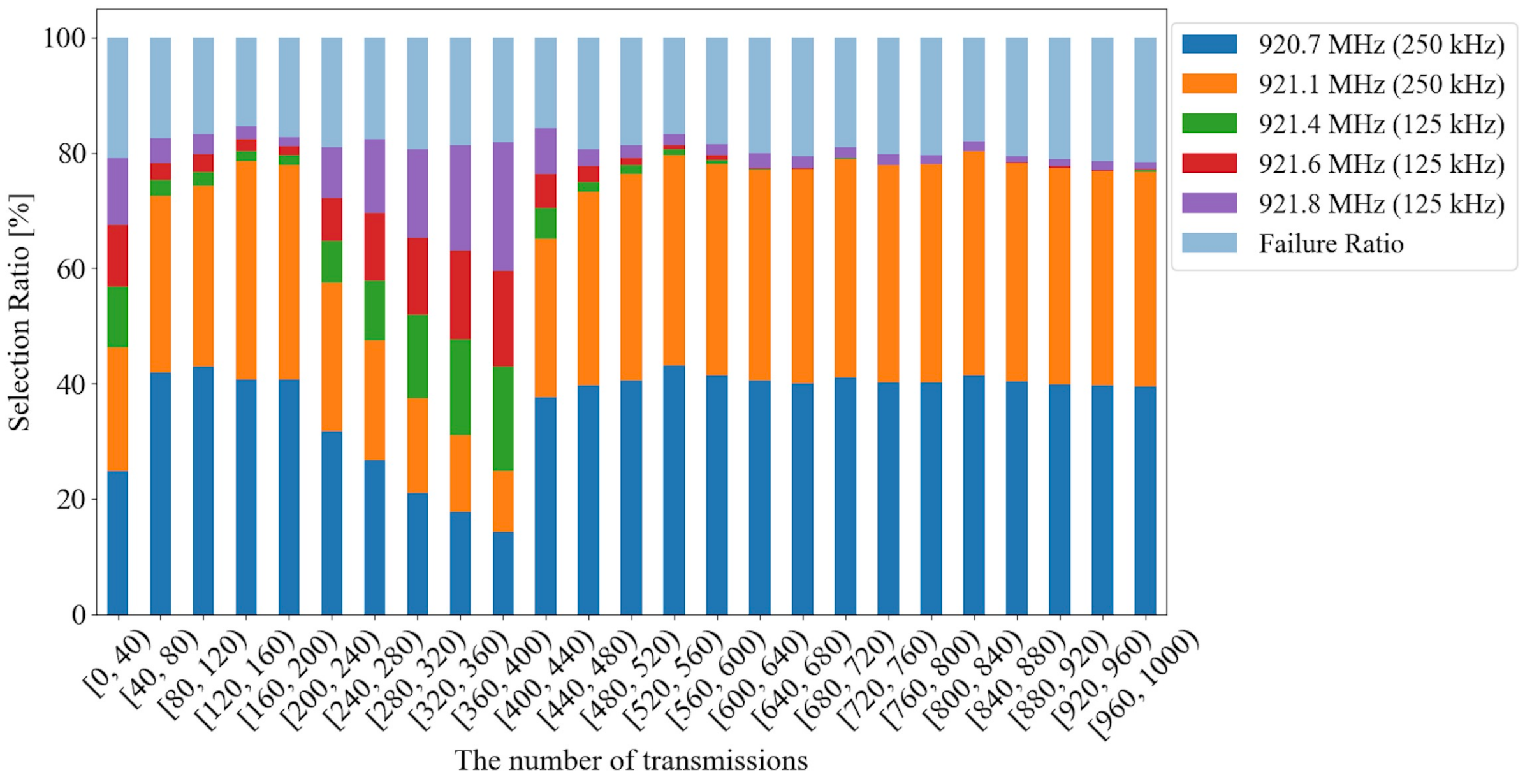}}
\caption{Selection ratio (UCB1-tuned)}
\label{SIC_1010}
\end{figure}

The overall transmission success rate over 1,000 transmissions was 76.98\% for the proposed method and 73.15\% for the baseline method. 
Fig.~\ref{SIC_SR} compares the variation in transmission success rate with the number of transmissions for both methods. 
As shown in Fig.~\ref{SIC_SR}, even when the communication environment changed, the proposed method maintained a success rate that was comparable to or higher than that of the baseline. 
In particular, during the 201--400 transmission interval, the success rate of the proposed method was approximately 10\% higher than that of the baseline, indicating that the proposed method could quickly detect environmental changes and relearn more suitable transmission parameters.

Figs.~\ref{SIC_0926} and \ref{SIC_1010} illustrate the proportions of selected parameter combinations within every 40-transmission interval for the proposed method (Fig.~\ref{SIC_0926}) and the baseline method (Fig.~\ref{SIC_1010}), respectively. 
As shown in the figures, when the communication environment degraded, the proposed method rapidly decreased the selection ratios corresponding to unavailable channels (920.7 MHz and 921.1 MHz in the 201–400 transmission interval, and 921.4 MHz and 921.6 MHz in the 601–800 interval), while increasing the proportions corresponding to available channels.
This behavior demonstrates that the integrated SIC mechanism can detect environmental changes within a small number of transmissions and promptly trigger relearning by resetting outdated learning histories.
In contrast, the baseline method tended to continue selecting unavailable channels even after the environmental change and gradually shifted toward available channels; however, this transition occurred more slowly compared with the proposed method.

\subsection{Energy efficiency}
\begin{figure}[t]
\centerline{\includegraphics[width=75mm]{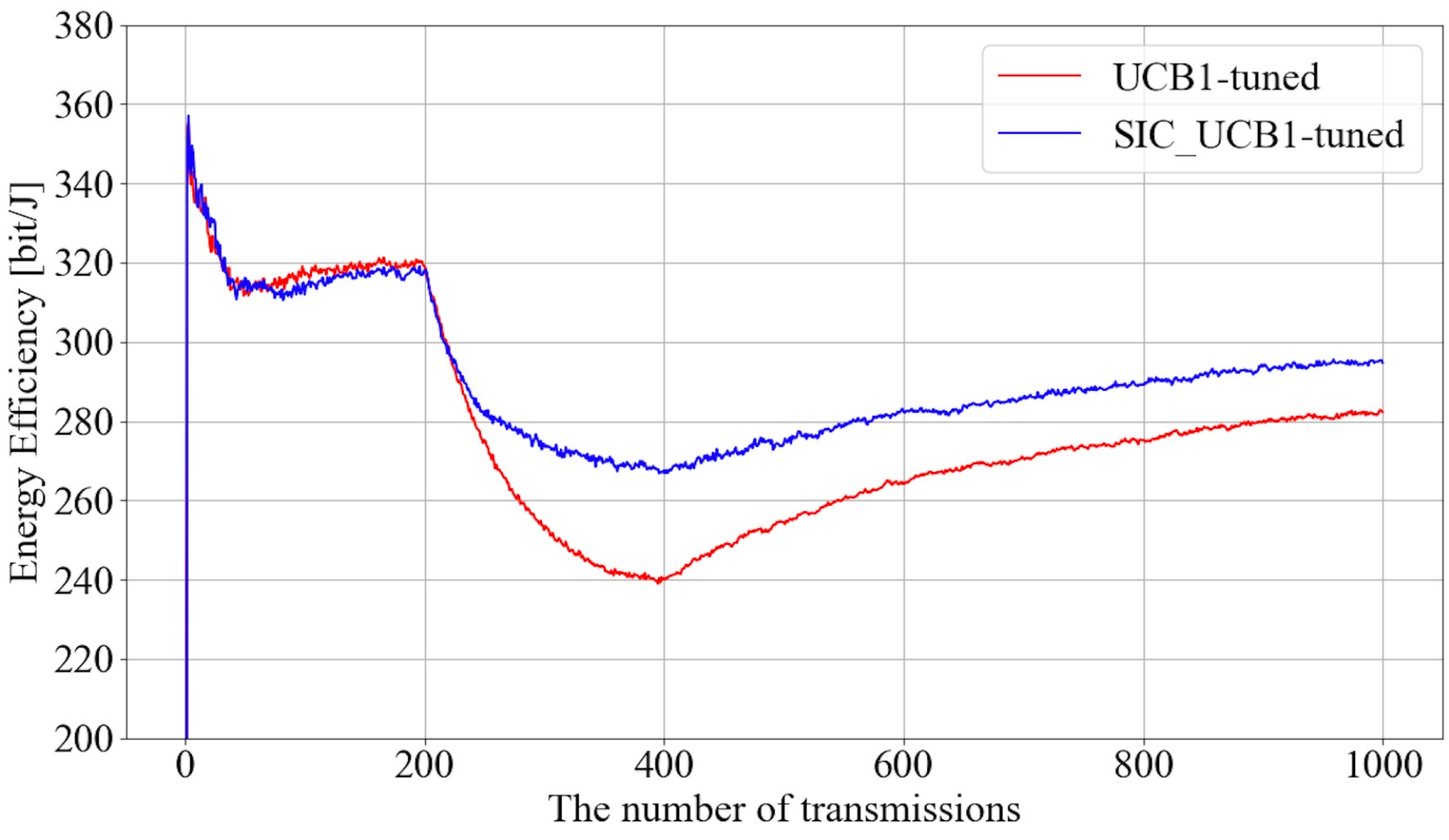}}
\caption{Energy efficiency versus the number of transmissions.}
\label{SIC_EE}
\end{figure}

The overall energy efficiency over 1,000 transmissions was 295.0 bit/J for the proposed method and 281.1 bit/J for the baseline method, confirming that the proposed method maintained higher long-term energy efficiency than the conventional approach. 
Fig.~\ref{SIC_EE} shows the transition of energy efficiency with respect to the number of transmissions. In the static communication environment before the 200th transmission, both the proposed method and the baseline method achieved a stable energy efficiency of approximately 320 bit/J. This indicates that under stationary conditions, both methods are capable of selecting transmission parameters that achieve high energy efficiency.

In contrast, when the communication environment changed after the 200th transmission and the 250 kHz channels became unavailable, both methods experienced a temporary degradation in energy efficiency. Because the baseline method relies on past learning history, it was slow to migrate toward the more energy-efficient 125 kHz channels, resulting in reduced transmission success rate and energy efficiency. In comparison, the proposed method detected the environmental change using SIC and reset its learning history, enabling rapid learning and selection of parameters suitable for the new environment. Consequently, by around the 400th transmission, the performance gap between the two methods widened to approximately 30 bit/J.
After the 400th transmission, when the 250 kHz channels became available again, the energy efficiency of both methods improved. In this phase, the baseline method also exhibited relatively high energy efficiency by leveraging the learning history accumulated in the initial environment. However, the energy loss caused by inefficient transmissions during the 200–400 interval remained as a cumulative penalty, and thus the proposed method still retains advantages in energy efficiency. 


\section{Conclusion}
\label{Conclusion}
In this paper, we proposed a SIC-aided MAB method that can adapt to dynamic changes in LoRa networks. Specifically, conventional MAB-based algorithms suffer from delayed adaptation when abrupt environmental changes occur, due to their reliance on past learning history. To address this issue, the proposed method incorporates SIC to statistically detect environmental changes and reset the learning history, enabling rapid re-adaptation.
Experimental results demonstrated that the proposed method outperforms the conventional UCB1-tuned algorithm in terms of both transmission success rate and energy efficiency. Specifically, it achieved a 3.8 \% improvement in success rate and a 4.9 \% enhancement in energy efficiency under the dynamic environment. Furthermore, it achieved faster adaptation after environmental changes.

\vspace{12pt}

\end{document}